\documentclass[aps,prl,twocolumn,superscriptaddress,showpacks]{revtex4-2}

\usepackage{comment}
\usepackage[colorlinks, citecolor=red]{hyperref}
\usepackage{hyperref}
\usepackage[utf8]{inputenc}
\usepackage[T1]{fontenc}    %
\usepackage[english]{babel}
\usepackage[pdftex]{graphicx}
\usepackage{amsmath,amssymb,amsthm,bbm, mathtools} %
\usepackage{mathrsfs}
\usepackage{tikz}   %
\usepackage[normalem]{ulem}

\begin{document}

\title{Andreev qubit readout from dynamic interference supercurrent}

\author{Xian-Peng Zhang}

\affiliation{Centre for Quantum Physics, Key Laboratory of Advanced Optoelectronic Quantum Architecture and Measurement (MOE), School of Physics, Beijing Institute of Technology, Beijing, 100081, China}

\affiliation{Department of Physics, Hong Kong University of Science and Technology, Clear Water Bay, Hong Kong, China}

\author{Chuanchang Zeng}

\affiliation{Centre for Quantum Physics, Key Laboratory of Advanced Optoelectronic Quantum Architecture and Measurement (MOE), School of Physics, Beijing Institute of Technology, Beijing, 100081, China}

\author{Zhen-Biao Yang}

\affiliation{Fujian Key Laboratory of Quantum Information and Quantum Optics,
College of Physics and Information Engineering,
Fuzhou University, Fuzhou,Fujian 350116, China}

\author{Jose Carlos Egues}
\email{egues@ifsc.usp.br}
\affiliation{Instituto de Fısica de S\~ ao Carlos, Universidade de Sao Paulo, 13560-970  S\~ ao Carlos, S\~ ao Paulo, Brazil}

\author{Yugui Yao}
\email{ygyao@bit.edu.cn}
\affiliation{Centre for Quantum Physics, Key Laboratory of Advanced Optoelectronic Quantum Architecture and Measurement (MOE), School of Physics, Beijing Institute of Technology, Beijing, 100081, China}

\begin{abstract}
Nondemolition protocols use ancilla qubits to identify the fragile quantum state of a qubit without destroying its encoded information, 
thus playing a crucial role in nondestructive quantum measurements particularly relevant for quantum error correction. 
However, the multitude of ancilla preparations, information transfers, and ancilla measurements in these protocols create 
an intrinsic overhead for information processing. Here we consider an Andreev qubit defined in a quantum-dot Josephson 
junction and show that the macroscopic time-dependent oscillatory supercurrent arising from the quantum interference of the 
many-body eigenstates, can be used to probe the qubit itself—arbitrarily close to the nondestructive limit—under currently available experimental capabilities. 
This readout of arbitrary superposition states of Andreev qubits avoids  ancillae altogether 
and significantly reduces experimental overhead as no repetitive qubit  reinitialization is needed. Our prediction of an AC-like Josephson effect \textit{without} an applied external voltage, which enables the nondestructive qubit readout, is a unique macroscopic manifestation of the microscopic dynamics of the Andreev quantum state. 
Our findings should have an unprecedented impact on advancing research and applications involving 
Andreev dots, thus positioning them as promising qubit contenders for quantum processing and technologies.
\end{abstract}

\maketitle

\textit{Introduction}
Andreev qubits~\cite{zazunov2003andreev,chtchelkatchev2003andreev}, combining the 
extraordinary scalability of superconducting circuits and the compact footprint of quantum dots, offer tremendous 
potential for advancements in quantum science and 
technology~\cite{bretheau2013exciting,willsch2024observation,pita2023direct,bernard2023long,bretheau2013supercurrent,zhang2024renormalized,janvier2015coherent,tosi2019spin,hays2021coherent,wendin2021coherent}. 
An exciting subject is the coherent manipulation of Andreev qubits within hybrid quantum electrodynamics architectures \cite{janvier2015coherent,tosi2019spin,canadas2021signatures,hays2020continuous,hays2021coherent,wendin2021coherent,metzger2021circuit}. 
In conventional quantum computation and information, the readout of a qubit state inherently collapses  the qubit state into one of 
the eigenstates of the measuring operator (e.g., the alive and dead  Schrödinger’s cat states) and thus generally destroys 
the previously prepared quantum state and its encoded information~\cite{nielsen2010quantum}. The qubit-state collapse during 
readout follows from the thought-to-be fundamental tenet of (von Neumann or projective) measurements in quantum mechanics,  
establishing the foundations for the no-cloning theorem~\cite{wootters1982single}.

\begin{figure}[h!]
\begin{center}
\includegraphics[width=0.98\linewidth]{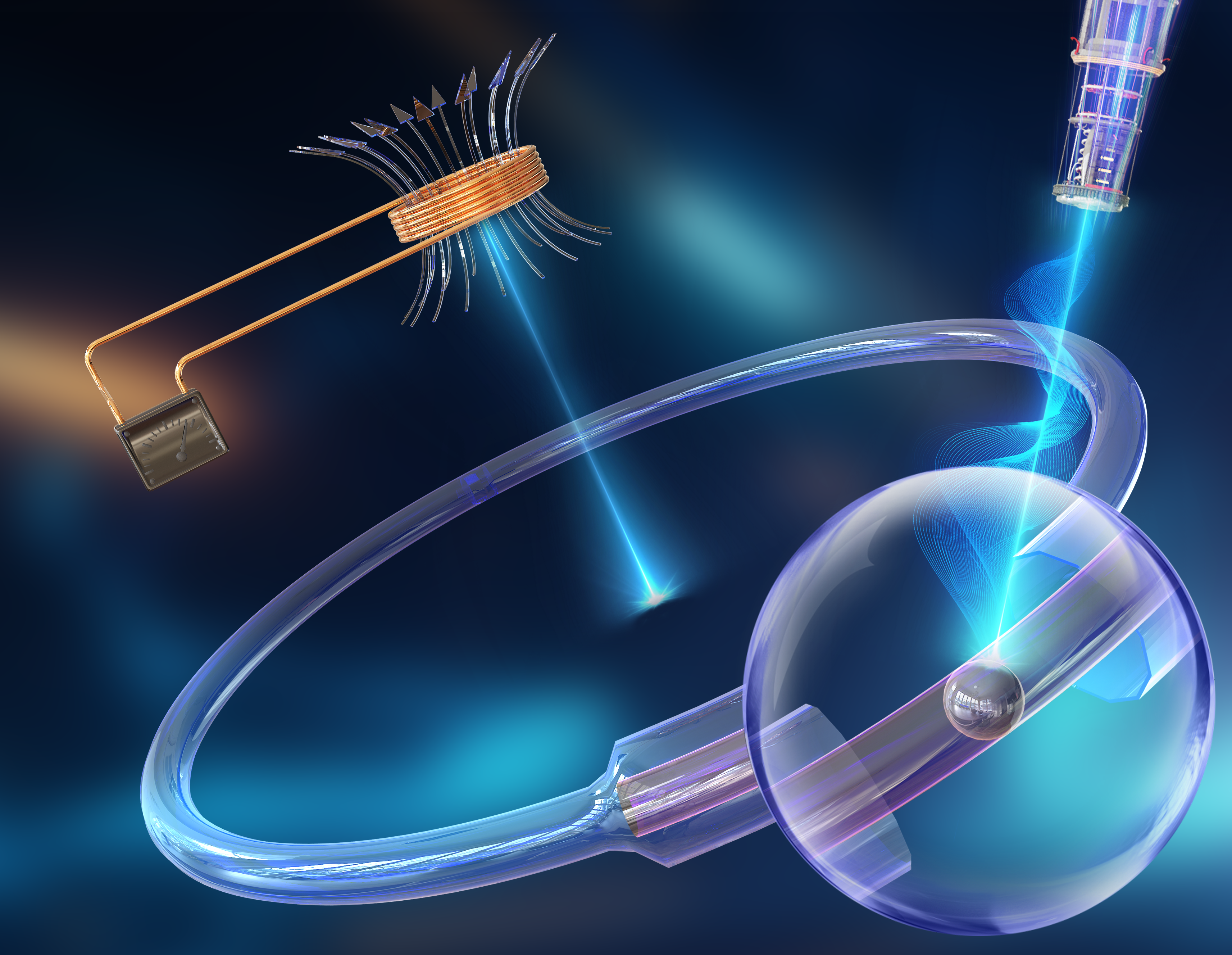}
\end{center}
\caption{Proposed experimental design for  measuring the supercurrent flowing through a quantum-dot Josephson junction (zoom) formed in a semiconducting nanowire (purple) and embedded in a large superconducting loop (blue). The small pickup coil (copper) of the SQUID magnetometer lies above the superconducting loop and is inductively coupled to the Andreev qubit (sphere in zoom). For a precise geometry and a detailed description of our proposed setup, see Fig. 1 in the  SM~\cite{SM}. }
\label{WSCar}
\end{figure}

To identify quantum states without destroying the encoded information, quantum non-demolition (QND) measurements  have been developed~\cite{braginsky1980quantum,braginsky1996quantum,xue2020repetitive,meunier2006nondestructive,nakajima2019quantum,wineland2013nobel,hume2007high,haroche2013nobel,nogues1999seeing,niemietz2021nondestructive,andersen2022quantum,jiang2009repetitive,neumann2010single,robledo2011high,maurer2012room,boss2017quantum,lupacscu2007quantum,johnson2010quantum,elder2020high}. These protocols  
rely on transferring and temporally storing the encoded qubit information to a well-prepared ancilla, which is then measured. 
Although the repetitive measurements of the ancillae can be destructive, the original qubit state remains intact  and retrievable. 
The ancilla-based repetitive QND readout has been a common approach across various platforms, such as trapped ions~\cite{wineland2013nobel,hume2007high}, superconducting spin qubits~\cite{hays2020continuous}, quantum-dot spin qubits~\cite{xue2020repetitive,nakajima2019quantum}, cavity quantum electrodynamics  systems~\cite{haroche2013nobel,nogues1999seeing,niemietz2021nondestructive,andersen2022quantum}, electron-nuclear spin qubits~\cite{jiang2009repetitive,neumann2010single,robledo2011high,maurer2012room}, and superconducting qubits~\cite{lupacscu2007quantum,johnson2010quantum,elder2020high}.
Even though these techniques have demonstrated significant progress, they introduce large-scale ancilla preparations, information transfers, and ancilla measurements~\cite{hume2007high} that are still expensive, laborious, and time-consuming in realistic 
experiments.

\begin{figure}[t!]
\begin{center}
\includegraphics[width=0.98\linewidth]{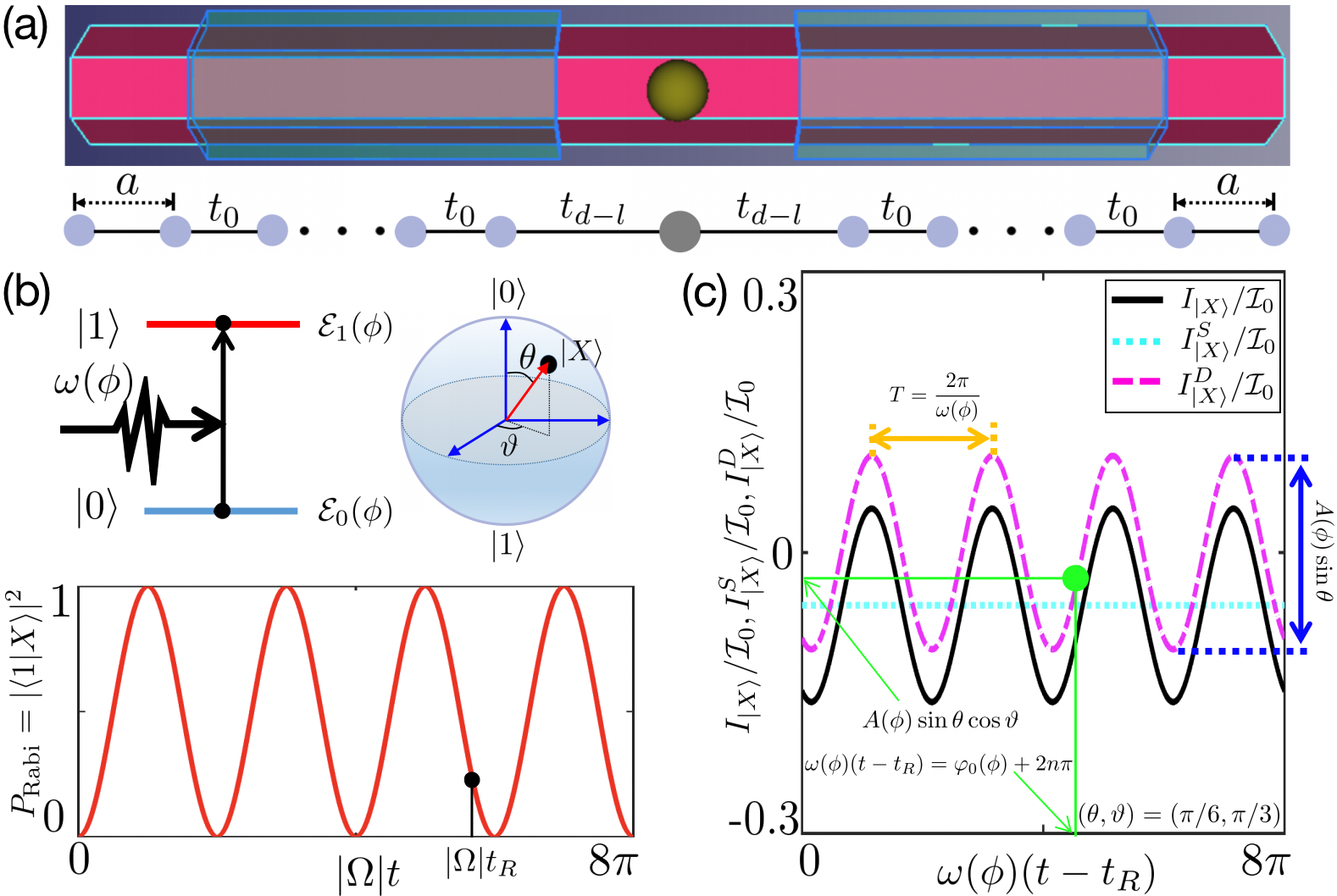}
\end{center}
\caption{(a) Sketch of a quantum-dot Josephson junction. A quantum dot (gray), gate-defined in an InAs nanowire (red), is tunnel coupled to proximitized InAs/Al left (L) and right (R) superconducting leads (blue) with phase bias $\phi=\phi_R-\phi_L$. The lower panel denotes  the corresponding 1D tight-binding model. (b) Preparation of an arbitrary Andreev qubit state: upon resonant microwave driving [$\hbar\omega(\phi)=\mathcal{E}_{1}(\phi)-\mathcal{E}_{0}(\phi)$], the qubit undergoes Rabi oscillations between states $\vert 0\rangle$ and $\vert 1\rangle$ with Rabi frequency  $\vert \Omega\vert $ [lower panel in (b)]. Arbitrary qubit states $\vert X\rangle =\cos(\theta/2)\vert 0 \rangle + \sin(\theta/2)e^{i\vartheta}\vert 1\rangle$ can be prepared on the Bloch sphere, where the azimuthal angle $\theta\in[0,\pi]$ is adjusted by the microwave pulse duration $t_R$ and the pole angle $\vartheta\in[0,2\pi)$ is controlled by the angle of the Rabi amplitude $\Omega$, i.e., $\vartheta=\text{angle}(\Omega)$. For instance, the black dot on the Rabi oscillation curve corresponds to $(\theta,\vartheta)=(\pi/6,\pi/3)$. (c) Supercurrent readout of Andreev qubits from the time evolution of the phase-biased SC $I^{}_{\vert X(t)\rangle}=I^{S}_{\vert X\rangle}+ A(\phi)\sin\theta\cos[\omega(\phi) (t-t_R)-\varphi_0(\phi)-\vartheta]$ corresponding to the arbitrary superposition 
$\vert X\rangle$ in (b), where $A(\phi)=2e\omega (\phi)\vert\mathcal{C}_{0,1}(\phi)\vert$, with $\mathcal{C}_{0,1}(\phi)=\langle 0 \vert   \partial_{ \phi}\vert 1\rangle=\vert\mathcal{C}_{0,1}(\phi)\vert e^{i\varphi_0(\phi)}$.  For a given experimental run, the phase bias $\phi$ and the gate-tunable dot level $\epsilon_D$ are kept fixed and so are $\omega(\phi)$, $\varphi_0(\phi)$, $\vert\mathcal{C}_{0,1}(\phi)\vert$,  $I_{\vert 0\rangle }$ and $I_{\vert 1\rangle}$. The static supercurrent $I^{S}_{\vert X\rangle}=\cos^2(\theta/2) I_{\vert 0\rangle}+ \sin^2(\theta/2) I_{\vert 1\rangle}$, calculated from the time average of the total SC,  determines $\theta$, which also controls the oscillation amplitude of the interference SC, i.e., $A(\phi)\sin\theta$. The qubit parameters, $\omega(\phi)$ and  $\vartheta$, can be read out from the oscillation period  $2\pi/\omega(\phi)$ and the interference SC  $A(\phi)\sin\theta\cos \vartheta$ at $\omega (\phi) (t-t_R)=\varphi_0(\phi)+2n\pi$, respectively. Other parameters are the same as Fig. \ref{WSC}.}  
\label{FIGSTORY} 
\end{figure}

\begin{figure}[t]
\begin{center}
\includegraphics[width=1.0\linewidth]{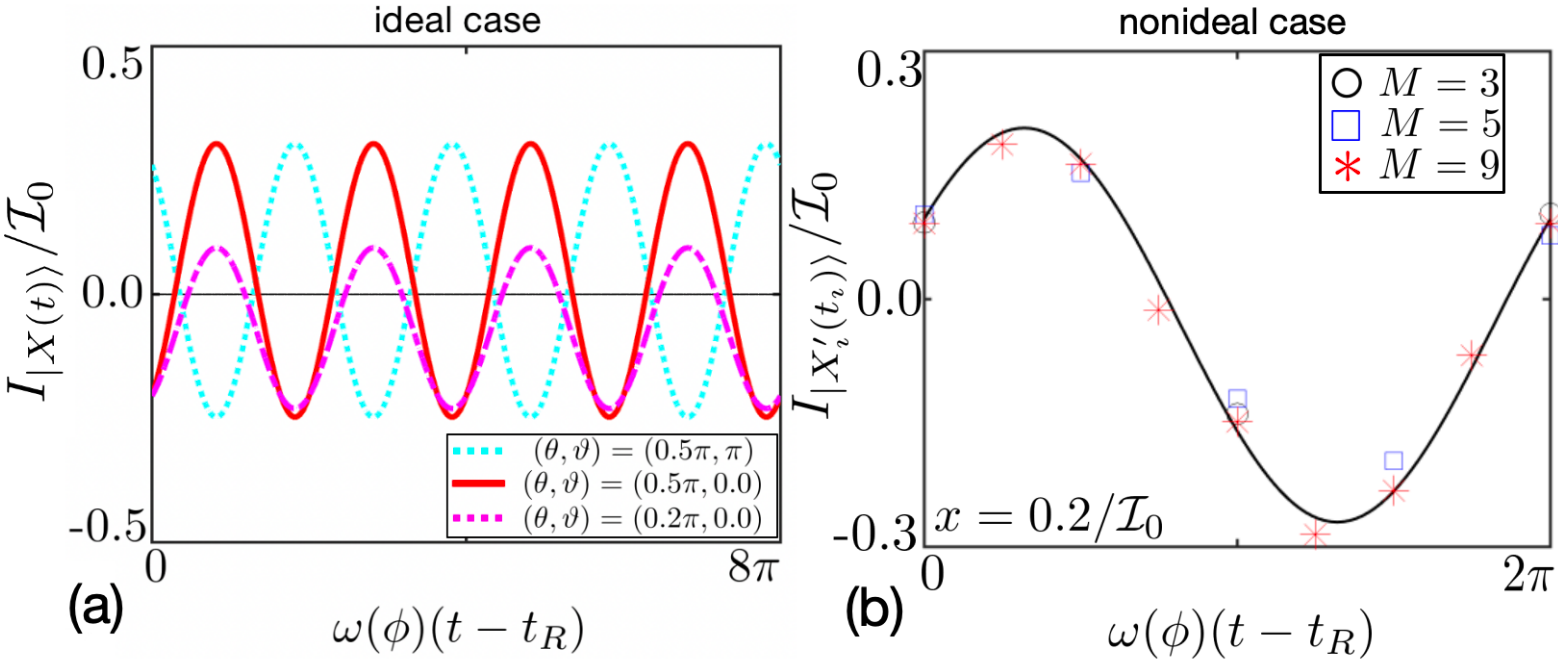}
\end{center}
\caption{(a) Ideal supercurrent $I_{\vert X (t)\rangle}$ readout for different initial qubit states $\vert X\rangle$, as parametrized 
by the Bloch-sphere angles $(\theta,\vartheta)$ [Fig.~\ref{FIGSTORY}(b)]. Both angles can be used to control the supercurrent, 
e.g., as $\vartheta$ changes from $0$ (red) to $\pi$ (cyan) (for $\theta=0.5\pi$), 
the supercurrent undergoes a $\pi$ shift, while by varying $\theta$ from $0.5\pi$ (red) to $0.2\pi$ (for ($\vartheta=0$) reduces its  
amplitude and modifies its static part [Eq.~\eqref{averagedsupercurrent}] from 
$I^S_{\vert X\rangle}/\mathcal{I}_0\simeq 0.028$ to $I^S_{\vert X\rangle}/\mathcal{I}_0\simeq-0.068$. 
(b) Real‑time supercurrent $I_{\vert X'_{\imath}(t_{\imath})\rangle}$[i.e., \eqref{vfdavfd}] under repeated quantum measurements for several values of the number of measurements $M$, where  $\sigma_I=0.02\mathcal{I}_0$ and $(\theta,\vartheta)=(\pi/3,\pi/2)$. 
Here we consider realistic parameters $\phi=1.18\pi$ and $\epsilon_D= 4\Delta$, resulting in 
$\hbar \omega(\phi)\simeq 0.89\Delta$, $\varphi_0(\phi)\simeq 0.82\pi$, $\vert  \mathcal{C}_{0,1}(\phi) \vert\simeq 0.31$, 
$I_{\vert 0\rangle}/\mathcal{I}_0=-0.091 $, and $I_{\vert 1\rangle}/\mathcal{I}_0=0.147$. Other parameters: $\Delta=1$, $t=t_R$, and $N=903$.  }
\label{WSC}
\end{figure}

In this work, we show that Andreev qubits formed in a quantum-dot Josephson junction [Fig.~\ref{FIGSTORY}(a)] can be used to overcome the shortcomings and overhead of QND readout. Our proposal benefits from an Andreev qubit originating a macroscopic quantum effect, namely, an intrinsic alternating supercurrent --similar to the AC Josephson effect but without any external voltage-- that is a unique fingerprint of its encoded quantum information.  Our envisioned Andreev qubit readout is enabled by an \textit{indirect} quantum measurement of the   
pickup-coil flux state of a superconducting quantum interference device (SQUID) magnetometer (Fig.~\ref{WSCar}) coupled to the Andreev qubit. It 
(i) arbitrarily approaches the nondestructive limit -- an inevitable shortcoming of \textit{direct} projective (von Neumann) quantum 
measurements on the qubit itself -- and (ii) obviates the extensive ancilla overhead (preparation, information transfer, measurements, etc), 
ubiquitous in state-of-the-art QND measurements. We also simulate the detrimental effects of experimental non-idealities in 
the current measurements, e.g., the inductive (random) back action of the small probing pickup coil on the Andreev qubit,  
and verify that our readout protocol has fidelities within the threshold for quantum error correction [Fig.~\ref{WSC}(b)].

\textit{Model and theory--}
We consider a typical quantum-dot Josephson junction~\cite{kurtossy2021andreev,vaitiekenas2021zero,wendin2021coherent,szombati2016josephson}, depicted in Fig.~\ref{FIGSTORY}(a), i.e., a quantum dot gate-defined in an InAs nanowire with dot energy $\epsilon_D$, tunnel-coupled to left ($j=L$) and right ($j=R$) proximitized InAs/Al leads with strength $t_{d-l}$. The leads have proximity-induced order parameters $\Delta e^{i\phi_j}$ with a flux-tunable phase difference $\phi = \phi_R - \phi_L$. Following the approach in  Ref.~\cite{zhang2024fabry}, we describe this hybrid system by an effective continuum model that maps onto the above tight-binding model with identical left and right leads having chemical potentials $\mu_S$ and nearest-neighbor tunneling amplitude $t_0$ and a discretization step $a$ [lower panel of Fig.~\ref{FIGSTORY}(a)]. This allows us to exactly diagonalize $H=\sum^{N}_{l=1}\sum_{s=\uparrow/\downarrow} E^{}_{ls}(\phi)\gamma_{ls}^{\dagger} (\phi)\gamma_{ls }^{}(\phi)+\mathcal{E}_G$, 
where $N$ represents the system size, and $s = \uparrow$, $\downarrow$ denotes the spin-up and spin-down Bogoliubons. 
The ground-state energy  $\mathcal{E}_{G}(\phi)=\mathcal{E}-\sum_{ls}\frac{1}{2}E^{}_{ls}(\phi)$ is $\phi$-dependent, with the 
constant $\mathcal{E}$ being $\phi$-independent.  The flux-tunable Bogoliubon operators associated with positive 
energies $E_{ls}(\phi)$ are represented as unit vectors in the Nambu space,
\begin{align} \label{fvakfkv}
    \gamma_{ls}(\phi)= \sum^{N}_{n=1}[(u_{s})^{}_{ln}c^{}_{ns}+(v_{s})^{}_{ln}(-sc^{\dagger}_{n-s})].
\end{align}
The field operator $c_{ns}^{}$ annihilates an electron with spin $s$ at site $n$. The $\phi$-dependent  $N\times N$ matrices $u_{s}$ and $v_{s}$ describe the electron and hole components of all Bogoliubons, respectively.
     
Quantum information processing based on Andreev qubits generally relies on the rigorous time evolution of the quantum-dot Josephson junction and thus demands precise many-body eigenfunctions, such as the ground-state eigenfunction.  The flux-tunable many-body ground state, defined by $\gamma_{ls}(\phi)\vert G\rangle=0$ for all $l$ and $s$ ($E_{ls}(\phi)>0$),  is full of Bogoliubov-like singlets
\begin{equation} \label{trufvdfmmvl}
    \vert G\rangle=\prod^{N}_{j=1}\frac{1}{N^{1/2}_{j}(\phi)}\left[1+\mathcal{A}^{}_{jj}(\phi)a^{\dagger}_{j\uparrow}(\phi)a^{\dagger}_{j\downarrow}(\phi)\right]\vert 0\rangle_e,
\end{equation}
where $N^{}_{j}(\phi)=1+\vert\mathcal{A}^{}_{jj}(\phi)\vert^2$ and $\vert 0\rangle_e$ is the electron vacuum. The superconducting pairs $a^{\dagger}_{j\uparrow}(\phi)a^{\dagger}_{j\downarrow}(\phi)$ show how the electrons in a non-uniform superconductor pair with each other (see expressions for Andreev coefficient $\mathcal{A}^{}_{jj}(\phi)$ and Andreev cloud $a^{\dagger}_{js}(\phi)$ in Ref.~\cite{zhang2024fermi}). Interestingly, the flux-tunable superconducting pairs carry an electrical signal, manifest via the macroscopic SC -- an inherent property of the superconducting ground state. 

Below, we present the exact time evolution of the system and analytically derive the supercurrent of an initial arbitrary superposition state.
Hereafter, we focus on the low-energy subgap (Andreev) space of the quantum-dot Josephson junction. This is a good approximation when the Andreev levels [$E_{ls}(\phi)$ with  $l=1$] are far away from the (quasi) continuum BdG spectrum [$E_{ls}(\phi)$ with $l> 1$]. Let us consider the Andreev level qubit as an example. Its even-parity logic states are represented by the two flux-tunable low-energy many-body eigenstates $\vert 0\rangle=\vert G \rangle$ and $\vert 1\rangle=\gamma^{\dagger}_{1\uparrow}(\phi)\gamma^{\dagger}_{1\downarrow}(\phi)\vert G \rangle$~\cite{zazunov2003andreev,janvier2015coherent,bretheau2013supercurrent,bretheau2013exciting}, having energies $\mathcal{E}_0(\phi)=\mathcal{E}_G(\phi)$ and $\mathcal{E}_1(\phi)=\mathcal{E}_G(\phi)+E_{1\uparrow}(\phi)+E_{1\downarrow}(\phi)$, respectively. In a hybrid architecture~\cite{janvier2015coherent,bretheau2013exciting}, the Andreev qubit coupling to microwave cavity photons allows one to prepare an arbitrary superposition $\vert X\rangle$ of the logic states, i.e., $\vert X\rangle =\cos(\theta/2)\vert 0 \rangle + \sin(\theta/2)e^{i\vartheta}\vert 1\rangle$ with $\theta\in[0,\pi]$ and $\vartheta\in[0,2\pi)$. These state vectors correspond to points ($\theta$, $\vartheta$) on the Bloch sphere, Fig.~\ref{FIGSTORY}(b).

\textit{Dynamic interference supercurrent--}
The supercurrent for the arbitrary superposition $\vert X\rangle$ parametrized 
by $(\theta,\vartheta)$ in Fig.~\ref{FIGSTORY}(b) is defined as the rate of change 
of the expectation value of  electron number operators in the left or right lead: 
$I_{\vert X\rangle}=+e \partial_t \langle X\vert N_L(t)\vert X\rangle =-e  \partial_t\langle X\vert N_R(t)\vert X\rangle$~\cite{cohen1962superconductive,ambegaokar1963tunneling} 
with $N_{j}(t)=\sum_{ns}c^{\dagger}_{jns}(t)c^{}_{jns}(t)$; here  
$c^{}_{jns}(t)=e^{+iHt/\hbar}c^{}_{jns}e^{-iHt/\hbar}$ (Heisenberg picture). This intrinsic supercurrent $I_{\vert X\rangle}$ flows  from the left lead into 
the right lead with no applied voltage. 
In contrast to 
previous results~\cite{bagwell1992suppression,beenakker1991universal,beenakker2013fermion}, 
our predicted supercurrent consists of a static part  $I^S_{\vert X\rangle}$ comprising the diagonal terms $e\cos^2(\theta/2)\partial_t\langle 0\vert N_j(t)\vert 0\rangle$ and 
$e\sin^2(\theta/2)\partial_t\langle 1\vert N_j(t)\vert 1\rangle$ plus a dynamic contribution $ I^{D}_{\vert X\rangle}(t)$ given by the two interference 
terms $(e/2)\sin\theta e^{+i\vartheta}\partial_t\langle 0\vert N_j(t)\vert 1\rangle$ 
and $(e/2)\sin\theta e^{-i\vartheta}\partial_t\langle 1\vert N_j(t)\vert 0\rangle$, 
i. e., $I_{\vert X(t)\rangle} = I^S_{\vert X\rangle} + I^{D}_{\vert X(t)\rangle}$. 
More precisely, $I^{S}_{\vert X\rangle}$ can be cast as a weighed contributions of the 
eigenstates in the superposition,
\begin{align} \label{averagedsupercurrent}
    I^{S}_{\vert X\rangle}=\cos^2(\theta/2) I_{\vert 0\rangle}+ \sin^2(\theta/2) I_{\vert 1\rangle},
\end{align}
where $I_{\vert m\rangle}=\mathcal{I}_0\frac{\partial}{\partial \phi} \mathcal{E}_m(\phi)$ ($\mathcal{I}_0=2e/\hbar$) 
with $H\vert m\rangle=\mathcal{E}_{m}(\phi)\vert m\rangle$, $m=0,1$. The 
dynamic part $ I^{D}_{\vert X\rangle}$ (see Sec.~II of the Supplementary Material (SM)~\cite{SM})  is due to the interference between the qubit eigenstates and reads 
\begin{equation} \label{fvaakfvk-recast}
I^{D}_{\vert X(t)\rangle}/\mathcal{I}_0= A(\phi)\sin\theta\cos[\omega(\phi) (t-t_R)-\varphi_0(\phi)-\vartheta],
\end{equation} 
 where $\omega(\phi)=[\mathcal{E}_1(\phi)-\mathcal{E}_{0}(\phi)]/\hbar$ is the qubit frequency, $t_R$ is the (Rabi) pulse 
duration, and $A (\phi)=2e\omega(\phi) \vert\mathcal{C}_{0,1}(\phi)\vert$, with $\mathcal{C}_{0,1}(\phi)=\vert\mathcal{C}_{0,1}(\phi)\vert e^{i\varphi_0(\phi)}$ 
denoting the matrix element $\mathcal{C}_{m,m'}(\phi)=\langle m\vert  \partial_ {\phi}\vert m'\rangle$ and $\varphi_0(\phi)$ its phase.  Note that $ I^{D}_{\vert X(t)\rangle}$ and $I^{S}_{\vert X\rangle}$ 
are of the same order, and the total supercurrent flows back and forth persistently, black curve in Fig.~\ref{FIGSTORY}(c).

Because $\vert X\rangle$ is a superposition of eigenstates, it evolves in time as $\vert X(t)\rangle =\cos(\theta/2)e^{-i\mathcal{E}_0(\phi)(t-t_R)/\hbar}\vert 0 \rangle + \sin(\theta/2)e^{i\vartheta-i\mathcal{E}_1(\phi)(t-t_R)/\hbar}\vert 1\rangle$ and the dynamic supercurrent \eqref{fvaakfvk-recast}  originates from the quantum interference between many-body eigenstates whose relative phase depends on time and qubit frequency. Consequently, the dynamic interference supercurrent exhibits an oscillatory pattern analogous to the usual alternating-current (AC) Josephson effect. This AC-like Josephson effect occurs in the absence of an applied voltage and is an interesting prediction in its own right. Interestingly, destructive interference emerges at degeneracy points where $\omega(\phi)=0$, leading to a vanishing dynamic supecurrent $ I^{D}_{\vert X(t)\rangle}=0$. At these degeneracies, any qubit superposition $\vert X\rangle$ becomes an eigenstate of the Andreev qubit, effectively reverting to the scenario with a purely static supercurrent. In contrast, for $\omega(\phi)\neq 0$, the 
supercurrent $ I^{D}_{\vert X(t)\rangle}$ depends crucially on the $\phi$-derivative of the many-body eigenstates via the matrix elements 
$\mathcal{C}_{m,m'}(\phi)$.  This dynamic interference supercurrent fundamentally distinguishes an Andreev qubit from others, as it highlights 
its quantum nature uniquely manifest macroscopically.

Figure~\ref{WSC}(a) shows $I_{\vert X(t)\rangle}$ for different values of $(\theta,\vartheta)$. The variation 
of $\vartheta$ from $0$ to $\pi$ results in a $\pi$ shift of the time oscillation. Additionally, the change of $\theta$ from $0.5\pi$ to $0.2\pi$ 
has two effects on the supercurrent. First, it reduces the amplitude of the time oscillation. Second, it changes the sign of the static SC 
from $I^S_{\vert X\rangle}/\mathcal{I}_0\simeq 0.028$  to $ I^S_{\vert X\rangle}/\mathcal{I}_0\simeq-0.068$. Hence,  
both $\vartheta$ and $\theta$ play essential roles in determining the behavior of the SC. More specifically, the Bloch 
 sphere angles $(\theta,\vartheta)$ can be straightforwardly obtained 
from $I_{\vert X (t)\rangle} $~vs.~$t$, which can be 
measured using a SQUID magnetometer (Fig.~\ref{WSCar}) as we discuss further below.  
From the static part $I^S_{\vert X\rangle }$ [Eq.~\eqref{averagedsupercurrent}], 
we can determine $\theta$ [$\theta$ also appears in the oscillation amplitude $A(\phi)\sin\theta$, Eq.~\ref{fvaakfvk-recast}]. 
On the other hand, the phase $\vartheta$ can be extracted from $I^D_{\vert X(t)\rangle}$ by identifying where  
$\omega(\phi) (t-t_R)=\varphi_0(\phi)+2n\pi$ [or $\omega(\phi) (t-t_R)=\varphi_0(\phi)+(4n+1)\pi/2$] in Fig.~\ref{FIGSTORY}(c), 
for which $I^{D}_{\vert X(t)\rangle}/\mathcal{I}_0=A(\phi)\sin\theta\cos\vartheta$ (cf. Eq.~\ref{fvaakfvk-recast})  [or $I^{D}_{\vert X(t)\rangle}/\mathcal{I}_0=A(\phi)\sin\theta\sin\vartheta$]. Next we present an experimental proposal for measurement of
Andreev qubits via $I_{\vert X(t)\rangle}$.

\textit{Indirect quantum measurement--}
Figure~\ref{WSCar} illustrates our envisioned experimental setup~\cite{braginsky1995quantum}: the Andreev qubit, embedded in
a large superconducting loop, is inductively coupled to a tiny pickup coil of a SQUID magnetometer (with  
sensitivities of fractions of the magnetic-flux quantum $\Phi_0$). Because of  the mutual inductance between the large loop and the pickup coil $L$, 
the quantum state $\vert \psi \rangle$ of the combined system is the entangled state 
$\vert \psi(t)\rangle=\sum_{n} [A^{n}_{\vert X(t)\rangle}\vert \Phi_{n} \rangle \otimes \vert X(t)\rangle+B^{n}_{\vert X(t)\rangle}\vert \Phi_{n} \rangle \otimes \vert \bar{X}(t)\rangle ]$, 
where $\vert \Phi_n \rangle$ denotes the flux eigenstate of the pickup coil and $\vert \bar X(t)\rangle$ is the orthogonal qubit state 
such that $\langle X (t)\vert \bar X(t)\rangle=0$. Here $A^{n}_{\vert X(t)\rangle} = \langle \Phi_n, X(t)\vert \psi(t) \rangle$ and 
$B^{n}_{\vert X(t)\rangle}= \langle \Phi_n, \bar X(t)\vert \psi(t) \rangle$ with $\vert \psi(t)\rangle$ chosen such  
that  $\vert \psi(t) \rangle \rightarrow \vert \psi(t)\rangle^{0}_{X}=\vert \Phi_n=0\rangle\otimes\vert X(t)\rangle$ when the large loop and the pickup coil are decoupled. 
The second term in $\vert \psi (t)\rangle$ describes the small admixture of the $\vert \bar X(t) \rangle$ qubit state to our qubit state $\vert X(t) \rangle$ due to the back action of the pickup coil.

Upon measuring the flux operator $\hat \Phi$ of the pickup coil, the probability of obtaining an eigenvalue $\Phi_{n}$ 
is $p^n_{\vert X(t)\rangle} = |A^{n}_{\vert X(t)\rangle}|^2 + |B^{n}_{\vert X(t)\rangle}|^2$. 
From the von Neumann projective postulate,  the state of the combined system immediately  
\textit{after} the pickup coil measurement  at an arbitrary time $t$, which collapses its quantum state into the measuring operator eigenstate 
$|\Phi_{n(t)}\rangle$, is  $\vert \psi'(t) \rangle= \vert \Phi_{n(t)} \rangle\otimes \vert X^\prime (t)\rangle$ with $\vert X^\prime (t)\rangle=(A^{n(t)}_{\vert X(t)\rangle}\vert X(t)\rangle+B^{n(t)}_{\vert X(t)\rangle}(t)\vert \bar{X}(t)\rangle )/\sqrt{p^{n(t)}_{\vert X(t)\rangle}}$. Here, both subscript and superscript $n(t)$ imply that the measured flux state of the pickup coil depends on the time-dependent $I_{\vert X\rangle} (t)$. 
Importantly, the \textit{indirect} quantum measurement of our Andreev qubit via the SQUID makes our qubit
state change slightly, due to the back action of the pickup coil. After the measurement, the combined system is ``naturally reset" into $\vert X^\prime (t)\rangle$  until the next  measurement. Thus, our setup enables one to uninterruptedly perform subequent quantum measurements at $t=t_0, t_1, t_2, t_3... $ as time continually evolves, 
without the need to reinitialize the qubit state.

We can quantify the back-action effect by estimating the additional phase difference produced in the Andreev qubit loop $\delta\phi_{n(t)}/2\pi=LI_{\text{coil}}(\Phi_{n(t)})/\Phi_0$, where $I_{\mathrm{coil}}(\Phi_{n(t)})$ is the supercurrent 
 in the pickup coil after the measurement; this is induced by the time-dependent flux $\Phi_{n(t)}$  threading it  (Lenz's law) due to $I_{\vert X\rangle }(t)$.  As a result, the combined system is described by the product state 
 $\vert \psi'(t)\rangle \simeq \vert \Phi_{n(t)} \rangle \otimes \vert X(\phi+\delta\phi_{n(t)})\rangle$, 
 i.e., $\vert X^\prime (t)\rangle\simeq \vert X(\phi+\delta\phi_{n(t)})\rangle \equiv A^{n(t)}_{\vert X(t)\rangle}(\delta\phi_{n(t)})\vert X(t)\rangle+B^{n(t)}_{\vert X(t)\rangle}(\delta\phi_{n(t)}) \vert \bar{X}(t)\rangle$. 
 For a negligible back-action ($\delta\phi_{n(t)}\to 0$), $|X'(t)\rangle \to |X(t)\rangle$ as $A^{n(t)}_{\vert X(t)\rangle}(\delta\phi_{n(t)}\to 0)\to 1$ and $B^{n(t)}_{\vert X(t)\rangle}(\delta\phi_{n(t)}\to 0)\to 0$. 
 This should be so because $I_{\text{coil}}(\Phi_{n(t)})\ll I_{\vert X\rangle}(t)$. 
 For example, for $I_{\mathrm{coil}}(\Phi_{n(t)})\simeq 0.1\mu\mathrm{A}$ ($\ll I_{\vert X\rangle} (t)\sim 1 \mu\mathrm{A} - 10  \mu\mathrm{A}$ \cite{golubov2004current,hsiang1980superconducting}), one finds $\delta\phi_{n(t)}/2\pi\sim 10^{-3}$~\cite{SM}. Thus, 
 the measurement back-action flux is negligible for small $I_{\mathrm{coil}}$, which, in principle, can enable the nondestructive 
 limit where $|X'(t)\rangle=|X(t)\rangle$.

Finally, we simulate the intrinsic effect of quantum measurement on the readout infidelity, which intrinsically originates from the stochastic collapse of flux states within the pickup coil. The state parameters $\theta$ and $\vartheta$ can, in principle, be extracted from just two time points of the supercurrent, i.e., $\{t_1,I_{\vert X(t_1)\rangle}\}$, and $ \{t_2,I_{\vert X(t_2)\rangle}\}$. Since the detected magnetic flux is proportional to the supercurrent, we \textit{ad hocly} incorporate the stochastic collapse of the flux state by adding Gaussian noise with standard deviation $\sigma_I$ to the supercurrent. The simulated measurement outcome at the $\imath$th time step is therefore generated as 
\begin{align} \label{vfdavfd}
    I_{\vert X'_{\imath}(t_{\imath})\rangle}=\text{normrnd}[I_{\vert X'_{\imath-1}(t_{\imath})\rangle} ,\sigma_I],
\end{align}
with $t_{\imath}=\imath\delta t+t_R$, where $\text{normrnd}[\mu,\sigma]$ is a random number of the normal distribution with mean value $\mu$ and standard deviation $\sigma$. The time between consecutive quantum measurements is chosen such that $\delta t=2\pi/[M\omega(\phi)]\sim 0.1$ ns, which is much shorter than the coherence time of the Andreev qubit~\cite{janvier2015coherent,hays2021coherent}, with $M$ being the number of measurements. Importantly, each stochastic collapse of flux states slightly alters the state of a qubit. Consequently, the state parameter $X^{'}_{\imath}=(\theta'_{\imath},\vartheta'_{\imath})$ for the $\imath$th flux detection is influenced by prior measurements, as is the supercurrent  $I_{\vert X'_{\imath-1}(t_{\imath})\rangle}$ (where the subscript denotes the state conditioned on the previous result). This encodes the measurement back-action, as detailed in our iterative simulations in SM~\cite{SM}. Note that the exact expressions for $A^{n(t)}_{\vert X(t)\rangle}$ and $B^{n(t)}_{\vert X(t)\rangle}$ are very complicated and hence we account for the feedback effect of the quantum measurement by phenomenologically introducing a parameter $x$ such that $A^{n(t_{\imath})}_{\vert X'_{\imath}(t_{\imath})\rangle}=\sqrt{1-x^2I_{\vert X'_{\imath}(t_{\imath})\rangle}}$ and $B^{n(t_{\imath})}_{\vert X'_{\imath}(t_{\imath})\rangle}=xI_{\vert X'_{\imath}(t_{\imath})\rangle}$. The later implies that the larger $I_{\vert X'_{\imath}(t_{\imath})\rangle}$ (or $\Phi_{n(t_{\imath})}$) corresponds to larger $I_{\mathrm{coil}}(\Phi_{n(t_{\imath})})$, which results in strong feedback effect of quantum measurement. 
Figure \ref{WSC} (b) displays the real‑time supercurrent under repeated quantum measurements for several values of $M$. To quantify the readout efficiency, we define the infidelity $\mathcal{F}(X,X')=1-\vert\langle X\vert X'\rangle\vert^2$, where $\vert X'\rangle$ is the measured state corresponding to a single prepared initial state $\vert X\rangle$. The state $\vert X'\rangle$ is extracted by fitting the time‑dependent supercurrent  $I_{\vert X'_{\imath}(t_{\imath})\rangle}$ to the functional form $I_{\vert X'(t)\rangle}=I^{S}_{\vert X'\rangle}+A\sin\theta'\cos[\omega(\phi) (t-t_R)-\varphi_0(\phi)-\vartheta']$. For a fixed coupling $x=0.2/\mathcal{I}_0$ [Fig. \ref{WSC}(b)], the simulated infidelities of $M=3,5,9$, are approximately  $4.8$\textperthousand, $1.5$\textperthousand, $1.9$\textperthousand, respectively. These values lie well within standard quantum error‑correction thresholds. See SM~\cite{SM} for a detailed discussion of indirect quantum measurements,  additional realistic estimates and errors. Therefore,  our interference‑based supercurrent detection via a tiny pickup coil is sufficiently nondestructive for coherent quantum state monitoring.

\textit{Conclusion--} 
We explicitly showed that the supercurrent of an Andreev qubit formed in a quantum dot Josephson junction embedded in a superconducting loop can have both static and dynamic components, depending on the qubit initial state. This predicted dynamic supercurrent is similar to the AC Josephson current, but \textit{without} an applied voltage. We proposed a protocol to measure an arbitrary Andreev qubit state via ordinary and standard supercurrent measurements. Our envisioned setup relies on 
performing \textit{indirect} quantum measurements on the system. This can be accomplished by measuring the flux state of a tiny SQUID pickup coil lying far above the Andreev qubit loop, which does not directly project the qubit state but provides the Bloch sphere qubit angles ($\theta$,$\vartheta$) non-destructively. Our readout offers several advantages over standard non-demolition readouts as it avoids the ancilla-qubit overhead (preparation, information transfer, measurements, etc) and eliminates the repetitive resettings of the same initially prepared qubit state for its detection in subsequent times. We showed that our protocol is robust against experimental non-idealities (e.g., back action of the pickup coil) and has fidelities within the quantum-error correction thresholds. Our findings should promote Andreev qubits as a viable platform for quantum technologies.

\textit{Acknowledgement.---} This work is supported by National Key R$\&$D Program of China (Grant Nos. 2020YFA0308800, 2021YFA1401500), 
the Hong Kong Research Grants Council (Grant Nos. 16306220, 16304523), the National Natural Science Foundation of 
China (Grant Nos. 12234003, 12321004, 12022416, 12475015, 11875108), the National Council for Scientific and 
Technological Development (Grant No. 301595/2022-4) and the São Paulo Research Foundation (Grant No. 2020/00841-9).

\end{document}